\documentclass[twoside]{ilcws07}
\usepackage[latin1]{inputenc}
\usepackage[dvips]{graphicx,epsfig,color}
\usepackage{wrapfig,rotating}
\usepackage{amssymb,amsmath,array}
\newcommand{\rtsth} {\sqrt{s_{\mathrm{th}}}}
\newcommand{\rtspk} {\sqrt{s_{\mathrm{pk}}}}
\newcommand{\Nth} {N_{\mathrm{th}}}
\newcommand{\Bth} {B_{\mathrm{th}}}
\newcommand{\Npk} {N_{\mathrm{pk}}}
\newcommand{\Bpk} {N_{\mathrm{pk}}}
\newcommand{\gev}{\,\, \mathrm{GeV}}
\newcommand{\HR}{\rule{1em}{0.2pt}}

\renewcommand{\figurename}{\bf Figure}
\renewcommand{\tablename}{\bf Table}
\newcommand{\mycaption}[1]{\caption{\sl #1}}
\begin{document}
\title{
 Precision Measurement of a Particle Mass at the Linear Collider }
\author{C. Milst\'ene$^1$, A. Freitas $^2$, M. Schmitt$^3$, A. Sopczak$^4$
\footnote{presented by A. Sopczak}
\vspace{.3cm}\\
1-Fermi National Laboratory- Batavia-Il-60510 -USA.\\
EMAIL Address: caroline@fnal.gov
\vspace{.1cm}\\
2-Institut f\"ur Theoretische Physik,Universit\"at Z\"urich, \\
Winterthurerstrasse 190, CH-8057,  Z\"urich, Switzerland. 
\vspace{.1cm}\\
3- Northwestern University, Evanston, USA. 
\vspace{.1cm}\\
4- Lancaster University, Lancaster LA1 4YB, United Kingdom.  
}

\maketitle
\begin{abstract}
 Precision measurement of the stop mass at the ILC is done in a method based on cross-sections 
 measurements at two different center-of-mass energies. This allows to minimize both the statistical 
 and systematic errors.  In the framework of the MSSM, a light stop, compatible with electro-weak baryogenesis, 
 is studied in its decay into a charm jet and neutralino, the Lightest Supersymmetric Particle(LSP), as a candidate 
 of dark matter. This takes place for a small stop-neutralino mass difference.
 
\end{abstract}
\section{Introduction}

In this study we aim at the minimisation of the systematic uncertainties and of the statistical error \cite{url}. 
This is achieved by using a method which allows to increase the precision in two ways. We deal with a ratio of 
cross-sections at two energy points. This takes care of the systematic uncertainties by cancelations and we choose 
one of the energies to be at the threshold were the sensitivity to mass is maximale.
We will show that even though we are dealing with more realistic data than in \cite{stop_dm}, we improve substantially 
the precision in the mass measurement.
As in \cite{stop_dm}, we are considering the MSSM with R Parity conservation 
and a scenario in which a light stop co-annihilates with the Lightest Supersymetric Particle 
(LSP), the neutralino, to produce the right amount of dark matter relic density,namely, within the 
experimental precision of WMAP and the Sloan digital sky survey \cite{wmap}. 
Together with a light Higgs, a light right-handed stop also supports electroweak baryogenesis.
Our data now include hadronization and fragmentation of the stop before its decay as well as
fragmentation of the charm of the decay. This provides a rather big smearing of the particles 
produced and together with gluon radiation increases the number of jets. We will use two different 
approaches. First we will optimize a set of sequential cuts as in \cite{stop_dm}, then we will be 
using a multi-variable optimization of the neural-network type, the Iterative Discrimination Analysis(IDA). 
We do take also advantage of the polarization since we deal with an almost right-handed stop as required 
for E.W. baryogenesis. This allows us to enhance the signal while getting rid of a big part of the background.

\section{Mass Precision Measurement:the Method}
\begin{figure}[tb]
\vspace{-2pt}
\centerline{\includegraphics[width=45mm]{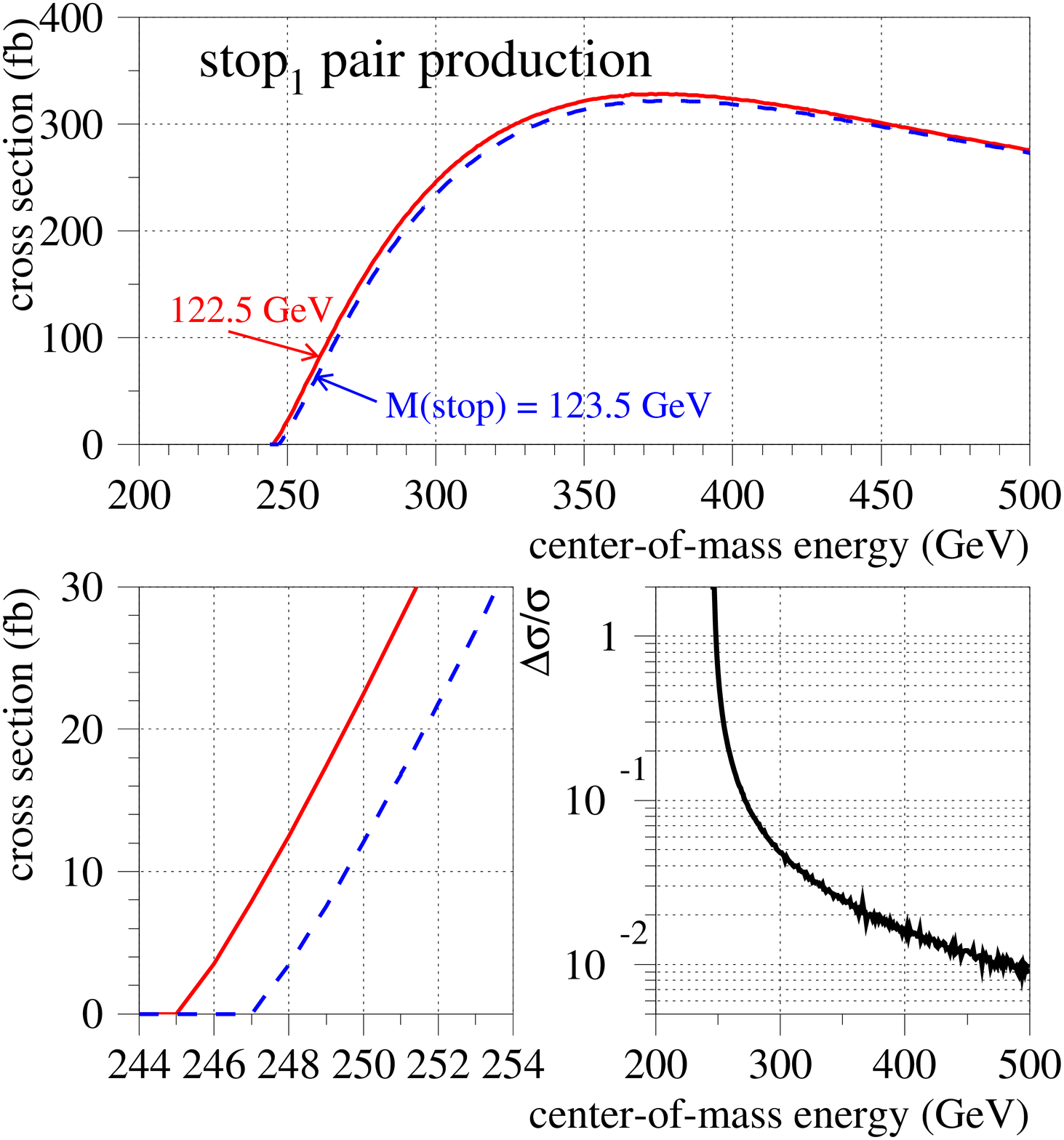}}
\caption{Precision in Pair Production Cross-Section}
\label{Fig:MV}
\end{figure}
  \begin{itemize}
  \item The production cross-section of stop pairs $e^+e^- \to \tilde{t}_1 \bar{\tilde{t}_1}$
  is represented to next to leading order (NLO), as a function of the energy for two hypothetical 
  values of the stop mass,122.5 GeV and 123.5 GeV, shown in Figure~\ref{Fig:MV}.
  \item In the lower left figure the scale has been blown up and one can see that the sensitivity 
  to small mass difference is high at or close to threshold while in the lower right figure one sees 
  that it is not the case at peak value.
  \item We will define a parameter Y, as a ratio of production cross-sections at two energy 
   points. This will reduce the systematic uncertainties in Y from the efficiencies as well 
   as from the beam luminosity measurements between the two energy points.     
 \item  One of the energy points is chosen at or close to the production energy
   threshold. This provides an increased sensitivity of Y to mass changes.
  \end{itemize}
 \vspace{-1pt}
\begin{figure}[hbt] 
\vspace{-20pt}
\centerline{\includegraphics[width=45mm, height=57mm]{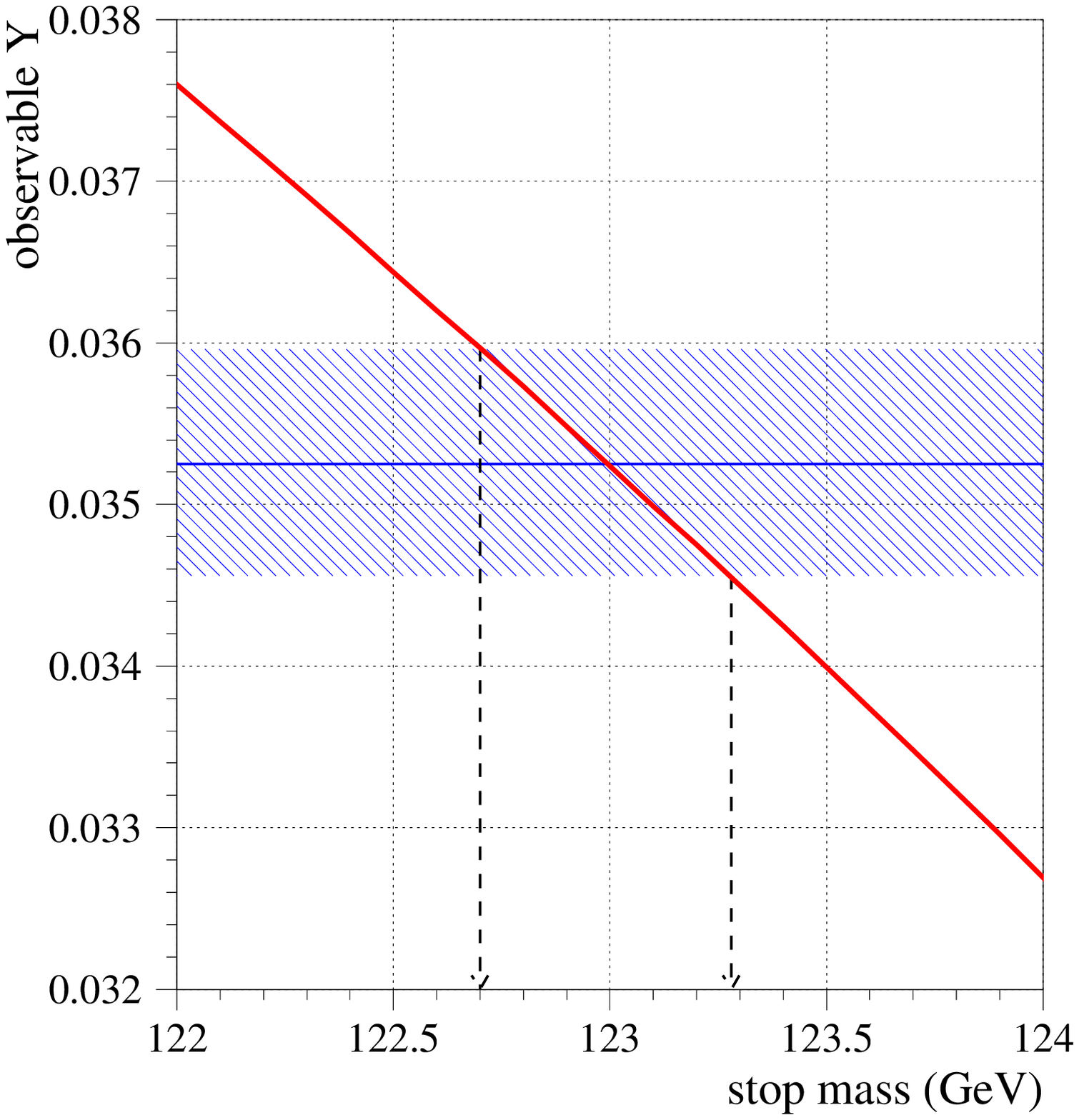}}
\caption{Precision in Determination of the Stop Mass}\label{Fig:YPAR}
\end{figure} 
\setlength\paperwidth  {210mm}
   \begin{equation}
     \label{def:Y}
     Y(M_X,\rtsth) \equiv
     \frac{\Nth-\Bth}{\Npk-\Bpk} = 
     \frac{\sigma(\rtsth) \epsilon_{th}L_{th}}{\sigma(\rtspk)\epsilon_{pk}L_{pk}} 
   \end{equation}  
$\sigma$ is the cross-section in [fb], N the number of detected data, B is the 
number of estimated background events, s is the square of the center of mass energy,
$\epsilon$ the total efficiency and acceptance and L is the integrated luminosity. 
The suffix (th) is used for the point at energy threshold and (pk) for the 
energy peak. M$_x$ is the mass to be determined with high precision.  
 
 As an example, one assumes 3\% precision for Y, The blue hashed region represents our measurements.
 One obtains a precision $\Delta$M$_x$ ~$\pm$0.2\%, the 2 vertical arrows. 

In the method, we determine the stop mass by comparing Y with the theoretical calculation of the cross-sections 
to next to the leading order (NLO) for both QCD and QED.It has been done for +80\% polarizations for the $e^-$ beam 
and $-60\%$ polarization for the $e^+$beam. 
 
\section{The Channel Studied $e^+e^- \to \tilde{t}_1 \bar{\tilde{t}_1}
\to cX^0 \bar{c}\bar{X^0}$}

A scan in the super-symmetry parameter space \cite{scan} has shown that a stop mass of 
122.5 GeV and a neutralino mass of 107.2 GeV are consistent with baryogenesis and dark matter.
The process and the background channels are listed below with their cross-sections with and
without polarization.
\renewcommand{\arraystretch}{1}%
\begin{table}[htb]
\vspace{-1pt}
\centering
\scalebox{.87}{
\begin{tabular}{|l|rrr|rrr|}
\hline
Process & \multicolumn{3}{c|}{Cross-section [pb] at $\sqrt{s} = 260 \gev$} 
       &  \multicolumn{3}{c|}{Cross-section [pb] at $\sqrt{s} = 500 \gev$} \\
\hline
$P(e^-) / P(e^+)$ & 0/0 & $-$80\%/+60\% & +80\%/$-$60\%
		  & 0/0 & $-$80\%/+60\% & +80\%/$-$60\% \\
\hline
$\tilde{t}_1 \tilde{t}_1^*$ & 0.032 & 0.017 & 0.077 & 0.118 & 0.072 & 0.276 \\
\hline
$W^+W^-$ & 16.9\phantom{0} & 48.6\phantom{0} & 1.77 
         & 8.6\phantom{0} & 24.5\phantom{0} & 0.77 \\
$ZZ$    & 1.12 & 2.28 & 0.99 & 0.49 & 1.02 & 0.44 \\
$W e\nu$ & 1.73 & 3.04 & 0.50 & 6.14 & 10.6\phantom{0} & 1.82 \\
$e e Z$  & 5.1\phantom{0} & 6.0\phantom{0} 
	 & 4.3\phantom{0} & 7.5\phantom{0} & 8.5\phantom{0} & 6.2\phantom{0} \\
$q \bar{q}$, $q \neq t$ & 49.5\phantom{0} & 92.7\phantom{0} & 53.1\phantom{0} 
			& 13.1\phantom{0} & 25.4\phantom{0} & 14.9\phantom{0} \\
$t \bar{t}$ & 0.0\phantom{0} & 0.0\phantom{0} & 0.0\phantom{0} & 0.55 & 1.13 &
0.50 \\
2-photon & 786\phantom{.00}&&&  936\phantom{.00}&& \\[-1ex]
$\quad p_{\rm t} > 5$ GeV &&&&&& \\
\hline
\end{tabular}}
\mycaption{The Cross-sections at $\sqrt{s}=260\ GeV$ and $\sqrt{s} = 500\ GeV$ 
for the signal and Standard Model background are given for different polarization 
combinations. The signal is given for a stop mixing angle of 0.01 and for a stop of 
$m_{\tilde{t}} =122.5\ Gev$,consistant with E.W. baryogenesis. The $e^-$ negative 
polarization values refer to left-handed polarization and positive values to 
right-handed polarization.}
\label{tab:xsec}
\end{table}
\renewcommand{\arraystretch}{1}%
\subsection{Simulations Characteristics}
The signal and background channels were generated with Pythia(6.129), the simulator
Simdet(4.03) and for the beamstrahlung Circe(1.0)\cite{Simdet}. They were generated in 
proportion with their cross-sections.  
\begin{itemize}
\item Hadronization of the $\tilde{t}_1$ quark and the fragmentation of the
charm quark come from the Lund string fragmentation model. We use Peterson
fragmentation \cite{peterson}. 
\item The stop Hadronization and fragmentation are simulated using T. Sjostrands code as described in detail 
by A.C.Kraan\cite{peterson}. The stop quark is set stable until after fragmentation, then 
it is allowed to decay. The stop fragmentation parameter is set relative to the bottom 
fragmentation parameter $\epsilon_{\tilde{t}} = \epsilon_b m_b^2
/ m_{\tilde{t}}^2$ and $\epsilon_b = -0.0050 \pm 0.0015$. Later improvements at LEP and a factor
two improvement assumed at ILC leads to  $\Delta \epsilon_t$=$0.6x10^-6$, as detailed in \cite{pub}.
The charm fragmentation is set from LEP to $\epsilon_c = -0.031 \pm 0.011$.
\item  The mean jet multiplicity increased for the data with fragmentation included.   
\end{itemize}  

\section{The Analysis}
The ntuple analysis code \cite{ntuple} which incorporates the Durham jet algorithm is used. The pre-selection 
and selection cuts are discribed in detail at both energies in \cite{pub}. 
For a right-handed stop, a run with right-chirality is favorable as shown in Table~\ref{tab:xsec}, one expects
at 500 GeV a luminosity of ~200 fb$^-1$ out of the 500 fb$^-1$. The luminosity of 200 fb$^-1$ is used for 
the 0/0 beam polarization as well as a point of comparison. 

 \subsection{The sequential cuts}
 Were made as similar as possible at the two energies to aim at the cancellation in Y of the systematics. The cuts and their 
 detailed results are given in \cite{pub}.In this analysis we allow two, three or four jets with the request that En<25 GeV
 for the lowest energy jets. The charm tagging is extracted using the ZVTOP software. The product of Charm tagging of the two 
 jets with the biggest charm probability is used to separate the signal from its main background, since the We$\nu$ has at most 
 one charm jet, whereas the signal has two charm-jets. 
 The backgrounds and the signal efficiencies are shown at the two energies after the sequential cuts in Table~\ref{tab:nev}.
\subsection{Iterative Discriminant Analysis (IDA)}
Combines the kinematic variables in parallel. The same kinematical variables and simulated 
events are used than in the cut-based analysis. A non-linear discriminant function followed by iterations 
enhances the separation signal-background.
\begin{figure}[h] 
\vspace{-2pt}
\centerline{\HR{\includegraphics[width=45mm, height=44mm]{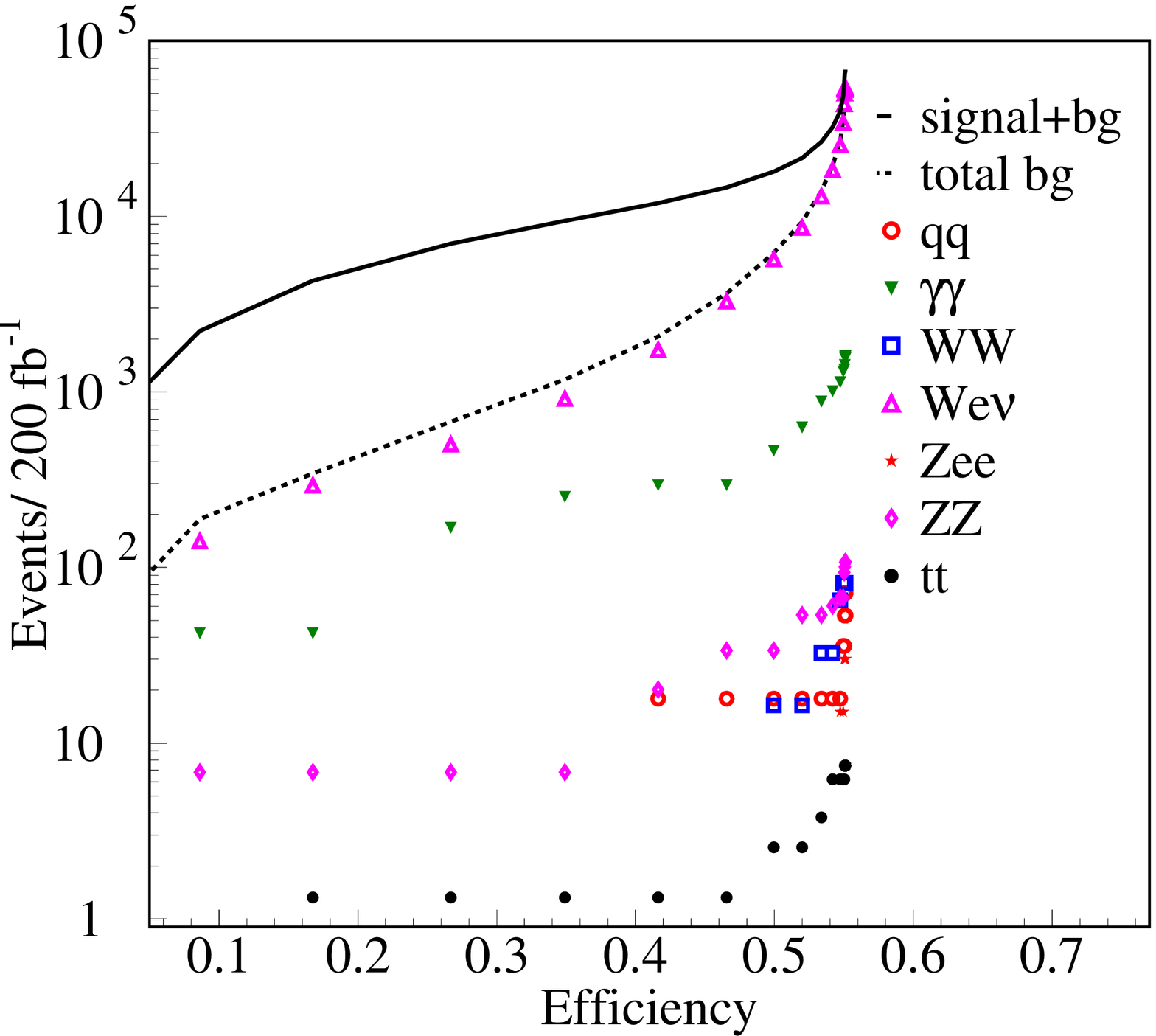}}\HR
\HR{\includegraphics[width=44mm, height=44mm]{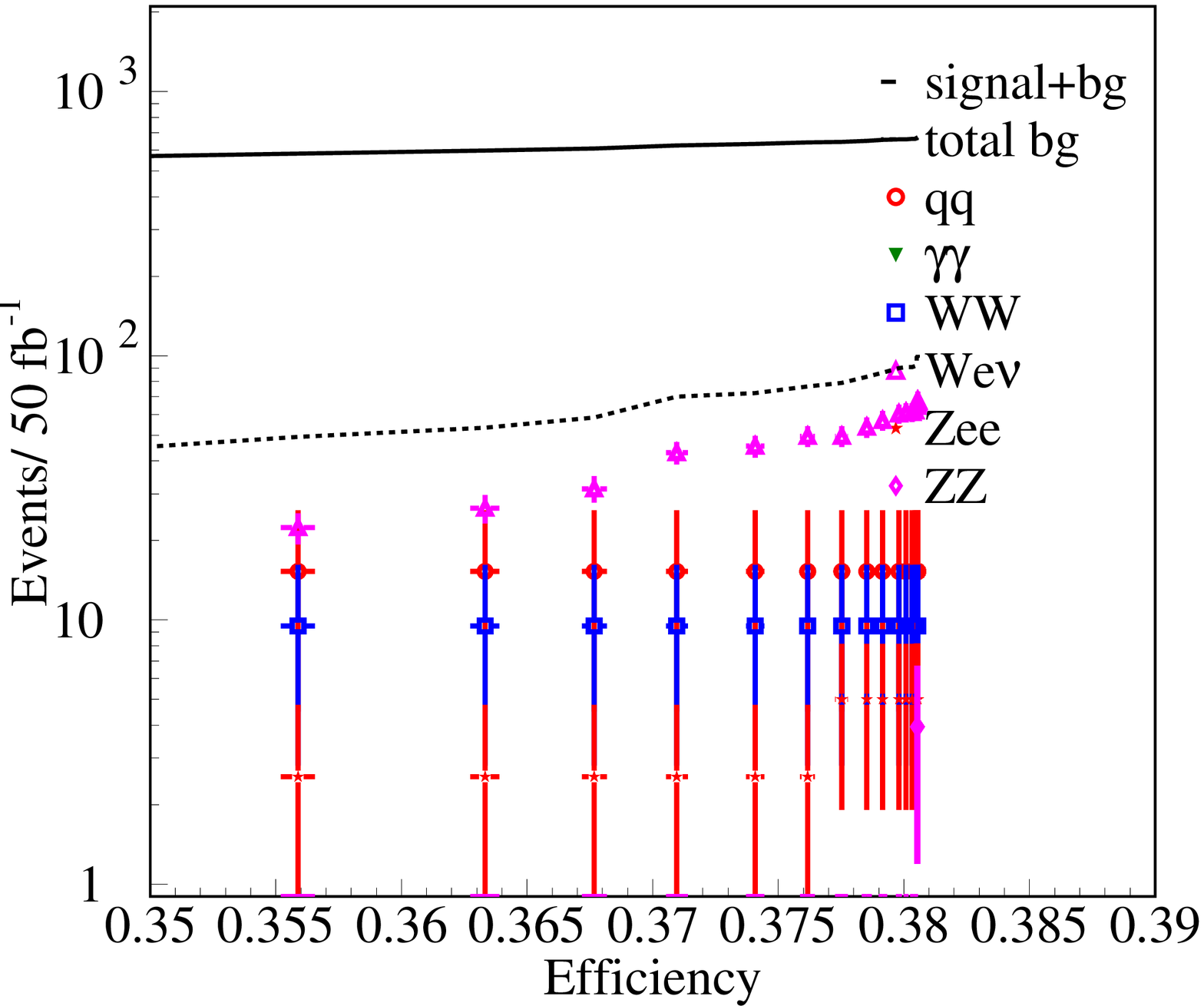}}\HR}
\caption{Detection Efficiency and Background Events at 500 GeV(left) and 260 GeV
(right).}\label{Fig:EFF}
\end{figure} 
 Both signal and background have been divided in equally sized samples, one used for the 
 training, the other as data. We will make two IDA iterations in our final analysis \cite{pub}. 
 The results are shown after a first IDA iteration for which one keeps 99.5\% of the signal
 efficiency followed by a second iteration. We assume the same luminosities and polarizarions 
 than for the sequential based analysis. The results are in Table~\ref{tab:nev}.
\renewcommand{\arraystretch}{1.}%
\begin{table}[hbt]
\vspace{1pt}
\centering
\scalebox{.85}{
\begin{tabular}{|l|r|rrl|r|rrl|rrl|rrl}
\hline
 &  \multicolumn{3}{c|}{${\cal L} = 50 \mbox{ fb}^{-1}$, at $260\gev$} 
 &  \multicolumn{3}{c|}{${\cal L} = 200 \mbox{ fb}^{-1}$,at $500\gev$}
 &  \multicolumn{2}{c|}{${\cal L} = 50 \mbox{fb}^{-1}$, at $260\gev$}
 &  \multicolumn{2}{c|}{${\cal L} = 200 \mbox{ fb}^{-1}$, at$500 \gev$} \\
\hline
 &  \multicolumn{3}{c|}{Sequential Cuts} 
 &  \multicolumn{3}{c|}{Sequential Cuts}
 &\multicolumn{2}{c|}{IDA }
 &\multicolumn{2}{c|}{IDA} \\
\hline
$Pe^- / Pe^+$ & generated &\multicolumn{2}{c|}{ 0/0 \ \ \  {Pol+}}
		  & generated &\multicolumn{2}{c|} {0/0 \ \ \ {Pol+}}
		  &\multicolumn{2}{c|}{0/0 \ \ \ {Pol+}}
		  &\multicolumn{2}{c|}{0/0 \ \ \ {Pol+}} \\
\hline
$\tilde{t}_1 \tilde{t}_1^*$ & 50,000 &\multicolumn{2}{c|}{544 \ \ 1309 } 
	& 50,000 &\multicolumn{2}{c|} {5170 \ \  12093} 
	&\multicolumn{2}{c|}{  619 \ \ \  1489}
	&\multicolumn{2}{c|}{9815 \ \  22958 }    \\
\hline
$W^+W^-$ & 180,000 &\multicolumn{2}{c|}{\ 38 \ \ \ \ \ \ 4} &210,000 &\multicolumn{2}{c|}{\ \ 16 \ \ \ \ \ \ \ \ 2} 
                  &\multicolumn{2}{c|}{ \ 11 \ \ \ \ \ \ \ \ 1} &\multicolumn{2}{c|}{\ \ $<$8 \ \ \ \ \ $<$ 1} \\
$ZZ$     & 30,000 &\multicolumn{2}{c|}{\ \ 8 \ \ \ \ \ \ \ 7 }& 30,000 &\multicolumn{2}{c|}{\ \ 36 \ \ \ \ \ \ \ 32}  
                  &\multicolumn{2}{c|}{\ \ $<$2\ \ \ \ \ \ \ $<$2}& \multicolumn{2}{c|}{\ \ 20 \ \ \ \ \ \ \  18}  \\
$W e\nu$ & 210,000 &\multicolumn{2}{c|}{ 208 \ \ \ \ \ 60} & 210,000 & \multicolumn{2}{c|}{7416 \ \ \  2198} 
                   &\multicolumn{2}{c|}{ 68 \ \ \ \ \ \ \  20} & \multicolumn{2}{c|}{1719 \ \ \ \ \ 510 }\\
$e e Z$  & 210,000 &\multicolumn{2}{c|}{\ \ 2\ \ \ \ \ \ \ 2} &  210,000 &\multicolumn{2}{c|}{\ \ \ $<$7 \ \ \ \ \ $<$6}  
                   &\multicolumn{2}{c|}{\ \ 3 \ \ \ \ \ \ \ \ 2 }&\multicolumn{2}{c|}{\ $<$7 \ \ \ \ \ \ \ $<$6}     \\ 
$q \bar{q}$, $q \neq t$ & 350,000 &\multicolumn{2}{c|}{\ 42 \ \ \ \ \ \ 45} & 350,000 & \multicolumn{2}{c|}{\ 15 \ \ \ \ \ \ \ 17}
&\multicolumn{2}{c|}{\ 16\ \ \ \ \ \ \ \ 17}&\multicolumn{2}{c|}{\ \ \ 18 \ \ \ \ \ \ \  21}  \\
$t \bar{t}$ &      &\multicolumn{2}{c|}{ - \ \ \ \ \ \ \  -}&  180,000 &\multicolumn{2}{c|}{\ \ 7 \ \ \ \ \ \ \ \ \ 7}
&\multicolumn{2}{c|}{- \ \ \ \ \ \ -}&\multicolumn{2}{c|}{\ \ \ \ \ 1 \ \ \ \ \ \ \  1 }    \\
2-photon & 1.6$\times 10^6$ &\multicolumn{2}{c|}{\ 53 \ \ \ \ \ \  53} & 
$8.5\times 10^6$ 
&\multicolumn{2}{c|}{\ 12 \ \ \ \ \ \ \ 12} &\multicolumn{2}{c|}{\ \ 27 \ \ \ \ \ \ \ 27}&\multicolumn{2}{c|}{\ \ 294\ \ \ \ \ 294}\\
\hline
\multicolumn{2}{c|}{Total Background}&\multicolumn{2}{c|}{\ \ 351\ \ \ \ \ 171}  
 &\multicolumn{3}{c|}{\ \ \ \ \ \ \ \ \ \ \ \ \ \ \ 7509 \ \ \ \ 2274}
 &\multicolumn{2}{c|}{127 \ \ \ \ \ \ \ 69}
 &\multicolumn{2}{c|}{\ 2067 \ \ \ \  851} \\
\hline
\multicolumn{2}{c|}{S/B}&\multicolumn{2}{c|}{\ \ \ \ \ \ 1.5 \ \ \ \ \ \ \ 7.6}  
 &\multicolumn{3}{c|}{\ \ \ \ \ \ \ \ \ \ \ \ \ \ \ \ \ \ \ \ \ 0.7\ \ \ \ \ \ \ 5.3}
 &\multicolumn{2}{c|}{\ \ \ 4.9\ \ \ \ \ \ 22}
 &\multicolumn{2}{c|}{\ \ \ \ \ 4.7 \ \ \ 27} \\
\hline
\multicolumn{2}{c|}{Efficiency}&\multicolumn{2}{c|}{ 0.340  }  
 &\multicolumn{3}{c|}{ 0.212 }
 &\multicolumn{2}{c|}{0.387}
 &\multicolumn{2}{c|}{0.416} \\
\hline
\end{tabular}}
\mycaption{Signal and background generated to NLO and after selection cuts are shown at
$\sqrt{s} = 260 \gev$ and $500 \gev$, for total luminosities of 50 fb$^{-1}$
and 200 fb$^{-1}$, respectively and the signal efficiencies.The event numbers after 
selection cuts are given without and with beam polarization. Pol+ = $Pe^- / Pe^+$
for $Pe^-$=+80\% and  $Pe^+$=-60\%.}
\label{tab:nev}
\end{table}
\renewcommand{\arraystretch}{1}%
The events $<$ N show the number of events corresponding to a single event.
\subsection{Contributions to the Mass Uncertainties}
 In Table~\ref{tab:sum} is summarised the contributions to the mass
 uncertainties
\renewcommand{\arraystretch}{1}%
\begin{table}[htb]
\vspace{-1pt}
\centering
\scalebox{.85}{
\begin{tabular}{|l|l|l|}
\hline
Error source for $Y$ & Cut-based analysis & Iterative Discriminant Analysis \\
\hline
Detector effects(systematics) & 0.9\% & 2.4\% \\
Charm  fragmentation (systematics) & 0.6\% & 0.5\%  \\
Stop fragmentation(systematics) & 0.7\% & 0.7\% \\
Neutralino Mass(systematics)    & 0.8\% & 2.2\% \\
Background Contribution(systematics) & 0.8\% & 0.1\% \\
Sum of experimental systematics & 1.7\%($\Delta m_{\tilde{t}_1}$=0.10 GeV) &
3.4\%($\Delta m_{\tilde{t}_1}$=0.21 GeV) \\
\hline
Statistical                     & 3.1\%($\Delta m_{\tilde{t}_1}$=0.19GeV) &
 2.7\%($\Delta m_{\tilde{t}_1}$=0.17GeV) \\ 
Sum of experimental errors & 3.5\%($\Delta m_{\tilde{t}_1}$=0.24 GeV) &
4.3\%($\Delta m_{\tilde{t}_1}$=0.28GeV) \\
\hline
Theory for signal cross-section & 5.5\% & 5.5\% \\
\hline
Total error $\Delta Y$ & 6.5\% ($\Delta m_{\tilde{t}_1}$=0.42 GeV)& 
7.0\%($\Delta m_{\tilde{t}_1}$=0.44 GeV) \\
\hline
\end{tabular}}
\mycaption{Combination of statistical and systematic errors for the
determination of the stop mass from a threshold-continuum cross-section
measurement. In parenthesis is given the overall error on the measured mass.
An beam spectrum error $\Delta m_{\tilde{t}_1}$=0.1 GeV,is included.}
\label{tab:sum}
\end{table}                                                               
\renewcommand{\arraystretch}{1}%


The next to next to leading order (NNLO) QCD corrections are expected to be of the 
same order than the NLO. This is based on the top quark results. 
Assuming a factor two improvement in the calculations by the time ILC is running
(A 1\% NNLO correction is also included for the EW componant). The relic dark matter
density is shown below 
\begin{figure}[hbt] 
\vspace{1pt}
\centerline{\includegraphics[width=58mm, height=60mm]{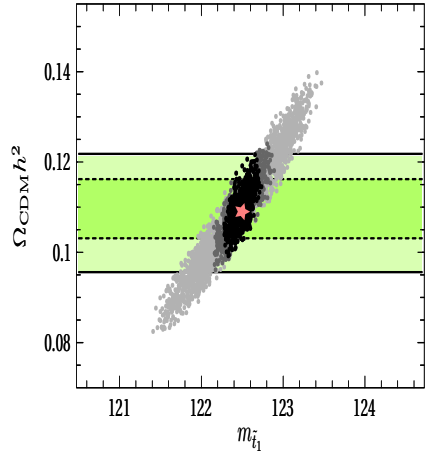}}
\caption{Dark matter Relic Density}
\label{Fig:OMEGA}
\end{figure} 
\section{Conclusions}
We  deal with more realistic data, including quarks hadronization and fragmentation, and with a lower integrated 
luminosity, almost by a factor two, but still manage to improve the stop mass precision by a factor three comparatively 
to \cite{stop_dm}.  
The results of the dark Matter relic density versus the stop mass precision are shown in three cases, in the last
figure. The light grey dots represent our previous results\cite{stop_dm} for $\Delta m_{\tilde{t}_1}$= 1.2 GeV, 
the dark grey dots correspond  $\Delta m_{\tilde{t}_1}$= 0.42 GeV, $\ \Omega_{CDM} h^2$ = 0.109+0.015-0.013,  
include both experimental and theoretical errors. The black dot imply $\delta m_{\tilde{t}_1}$= 0.24\ GeV\, .  
 $\Omega_{CDM} h^2$ = 0.109+0.0012-0.0010, experimental errors sequential cuts. It is only a small 
 improvement in the precision of the dark matter density with respect to $\Delta m_{\tilde{t}_1}$=0.42.
  The red star is our working point. The precision is very comparable to 0.103$<\Omega_{CDM}h^2<$0.116, the current 
 WMAP results are shown by the horizontal green bands on the figure for 1$\sigma$ and 2$\sigma$ constraints. 


\begin{footnotesize}

\end{footnotesize}


\end{document}